\documentclass[twocolumn,prl,showpacs,floatfix]{revtex4}

\usepackage{graphicx}
\usepackage{dcolumn}
\usepackage{bm}
\usepackage{amsmath}

\begin{document}

\title{
Crystalline boson phases in harmonic traps:\\
Beyond the Gross-Pitaevskii mean field
}

\author{Igor Romanovsky}
\author{Constantine Yannouleas}
\author{Uzi Landman}

\affiliation{School of Physics, Georgia Institute of Technology,
             Atlanta, Georgia 30332-0430}

\date{25 June 2004; Physical Review Letters {\bf 93}, 230405 (2004)}         

\begin{abstract}
Strongly repelling bosons in two-dimensional harmonic traps are 
described through breaking of rotational symmetry at the Hartree-Fock 
level and subsequent symmetry restoration via projection techniques, thus
incorporating correlations beyond the Gross-Pitaevskii (GP) solution. 
The bosons localize and form polygonal-ring-like crystalline 
patterns, both for a repulsive contact potential and a Coulomb interaction,
as revealed via conditional-probability-distribution analysis. 
For neutral bosons, the total energy 
of the crystalline phase saturates in contrast to the GP solution, and
its spatial extent becomes smaller than that of the GP condensate.
For charged bosons, the total energy and dimensions approach the values of
classical point-like charges in their equilibrium configuration.
\end{abstract}

\pacs{03.75.Hh, 03.75.Nt}

\maketitle

Bose-Einstein condensates (BEC's) in harmonic traps \cite{corn}
are normally associated with weakly interacting neutral atoms, and their physics
is described adequately by the Gross-Pitaevskii (GP) mean-field theory 
\cite{dal}. Lately, however, experimental advances in controlling the 
interaction strength \cite{cor,grei,par,wei} permit the production of novel 
bosonic states in the regime of strong interparticle repulsions. 
Theoretical efforts motivated by this capability include studies of the 
Bose-Hubbard model \cite{jak,mot}, and investigations about the 
``fermionization'' limit of an one-dimensional (1D) gas of trapped impenetrable
bosons \cite{gir2,dunj,ast}, often referred to as the Tonks-Girardeau 
(TG) regime \cite{gir2,gir}. In this Letter, we address the still open problem 
of strongly repelling (impenetrable) bosons in higher dimensions, and in 
particular in two dimensions (2D).

We describe the strongly repelling bosons through symmetry breaking at the
Hartree-Fock (HF) mean-field level 
followed by post-Hartree-Fock symmetry restoration, thus incorporating
correlations beyond the GP solution. This two-step method, which has not been
applied yet to the  bosonic many-body problem, has been shown to
successfully describe strongly correlated electrons in 
2D semiconductor quantum dots \cite{yl}. We focus here on results 
for 2D interacting bosons in a harmonic trap, with the extension to 3D 
systems being straightforward. 

To illustrate our method, we consider systems with a few bosons. 
The method describes the transition from a BEC state to a
crystalline phase, in which the trapped localized bosons form 
crystalline patterns. In 2D, these patterns are 
ring-like, both for a repulsive contact 
and a Coulomb interaction. At the mean-field level,
these crystallites are static and are portrayed directly in the
single-particle densities. After restoration of symmetry, 
the single-particle densities are rotationally symmetric, and thus the
crystalline symmetry becomes ``hidden''; however, it can be revealed in the
conditional probability distribution (CPD, anisotropic pair correlation), 
$P({\bf r},{\bf r}_0)$, which expresses the probability of finding a particle 
at ${\bf r}$ given that the ``observer'' (i.e., reference point) is riding on 
another particle at ${\bf r}_0$ \cite{yl4}. 

Mean-field symmetry breaking for bosonic systems has been discussed earlier in 
the context of two-component condensates, where each species is associated with 
a different space orbital \cite{esr}. We consider here one species of bosons,
but allow each particle to occupy a different space orbital $\phi_i({\bf r}_i)$.
The permanent
$|\Phi_N \rangle = {\it Perm}[\phi_1({\bf r}_1), ..., \phi_N({\bf r}_N)]$
serves as the many-body wave function of the {\it unrestricted\/}
Bose-Hartree-Fock (UBHF) approximation. This wave function
reduces to the Gross-Pitaevskii form with the {\it restriction\/}
that all bosons occupy the same orbital $\phi_0({\bf r})$,
i.e., $|\Phi^{\text{GP}}_N \rangle =\prod_{i=1}^N \phi_0({\bf r}_i)$, and
$\phi_0({\bf r})$ is determined self-consistently at the restricted 
Bose-Hartree-Fock (RBHF) level \cite{note1} via the equation \cite{esr2}
$ [ H_0({\bf r}_1) + (N-1) \int d{\bf r}_2 \phi^*_0({\bf r}_2)
V({\bf r}_1,{\bf r}_2) \phi_0({\bf r}_2)] 
\phi_0({\bf r}_1) = \varepsilon_0 \phi_0({\bf r}_1)$.
Here $V({\bf r}_1,{\bf r}_2)$ is the two-body repulsive interaction, 
which can be either a long-range Coulomb force, 
$V_C=Z^2e^2/(\kappa |{\bf r}_1 -{\bf r}_2|)$, for charged bosons
or a contact potential, $V_{\delta}= g\delta({\bf r}_1 -{\bf r}_2)$,
for neutral bosons.
The single-particle hamiltonian is given by $H_0({\bf r}) = 
-\hbar^2 \nabla^2 /(2m) + m \omega_0^2 {\bf r}^2/2$, where $\omega_0$ 
characterizes the harmonic confinement.

{\it First step: Symmetry breaking.\/}
Going beyond the GP approach to the {\it unrestricted\/} Hartree-Fock level
(i.e., using the permanent $|\Phi_N \rangle$) results in a set of UBHF equations
with a higher complexity than that encountered in electronic structure problems 
[13(a)]. Consequently, we simplify
the UBHF problem by considering explicit analytic expressions for the space
orbitals $\phi_i({\bf r}_i)$. In particular, since {\it the bosons must avoid 
occupying the same position in space in order to minimize their mutual
repulsion\/}, we take all the orbitals to be of the form of displaced 
Gaussians \cite{note2}, namely, $\phi_i({\bf r}_i) = \pi^{-1/2} \sigma^{-1} 
\exp[-({\bf r}_i - {\bf a}_i)^2/(2 \sigma^2)]$. The positions ${\bf a}_i$  
describe the vertices of concentric regular polygons, with both the width
$\sigma$ and the radius $a=|{\bf a}_i|$ of the regular polygons  
determined variationally through minimization of the total energy
$E_{\text{UBHF}} = \langle \Phi_N | H | \Phi_N \rangle$
/$\langle \Phi_N | \Phi_N \rangle$, where 
$H = \sum_{i=1}^N H_0({\bf r}_i) + \sum_{i < j}^{N} 
V( {\bf r}_i,{\bf r}_j)$ is the many-body hamiltonian.
 
With the above choice of localized orbitals, the unrestricted permanent 
$|\Phi_N \rangle$ breaks the continuous rotational symmetry. However,
for both the cases of a contact potential and a Coulomb force, 
the resulting energy gain becomes substantial for stronger repulsion.
Controlling this energy gain
(the strength of correlations) is the ratio $R$ between the strength of the 
repulsive potential and the zero-point kinetic energy. Specifically, for a 2D 
trap, one has $R_{\delta} = gm/(2\pi\hbar^2)$ for a contact potential and 
$R_W=Z^2e^2/(\hbar \omega_0 l_0)$ for a Coulomb force, with 
$l_0=\sqrt{\hbar/(m\omega_0)}$ being the characteristic harmonic-oscillator 
length. (The subscript $W$ in the case of a Coulomb force stands for 
``Wigner'', since the Coulomb crystallites in harmonic traps are finite-size 
analogs of the bulk Wigner crystal \cite{wig}.)

\begin{figure}[t]
\centering\includegraphics[width=6.2cm]{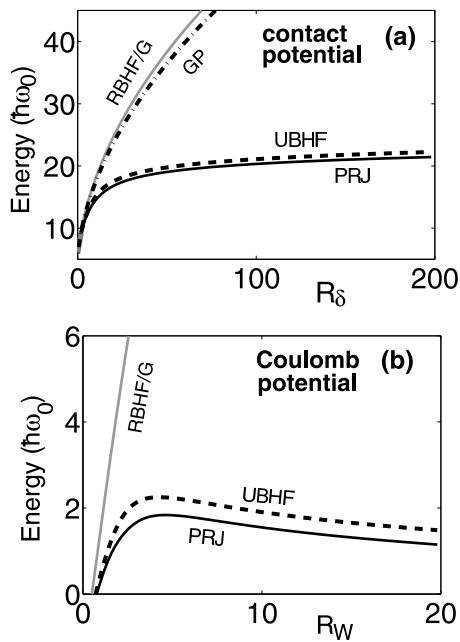}
\caption{
Total energies as a function (a) of $R_{\delta}$ and (b) of $R_W$ 
for various approximation levels, calculated for $N=6$ harmonically confined 
2D bosons in the (1,5) lowest-energy configuration.
Notation: RBHF/G - Restricted Bose-Hartree-Fock (RBHF) energy,
 $E^{Gauss}_{RBHF}$ , with the common 
orbital $\phi_0({\bf r})$ approximated by a Gaussian centered at the 
trap origin; GP - the Gross-Pitaevskii energy; PRJ - the energy of the
symmetry-restored state obtained via projection of the (unrestricted)
UBHF state. Energies in units of $\hbar \omega_0$.}
\end{figure}

In Fig.\ 1, we display  as a function of the
parameters $R_{\delta}$ (a) and $R_W$ (b), respectively, 
the total energies for $N=6$ bosons
calculated at several levels of approximation. 
In both cases the lowest UBHF energies 
correspond to a (1,5) crystalline configuration, namely one boson is at 
the center and the rest form a regular pentagon of radius $a$. Observe
that the GP total energies are slightly lower than the 
$E_{\text{RBHF}}^{\text{Gauss}}$ ones; however, 
both exhibit
an unphysical behavior since they diverge as $R_{\delta} \rightarrow \infty$.
This behavior contrasts sharply with that of the unrestricted Hartree-Fock
energies, $E_{\text{UBHF}}$, and those of the projected (PRJ)
states (see below), which saturate as 
$R_{\delta} \rightarrow \infty$; in fact, a value close to saturation is
achieved already for $R_{\delta}$ ($R_W$) $\sim$ 10. 
We have checked that for all cases with 
$N=2 - 7$, the total energies exhibit a similar behavior. For a repulsive 
contact potential, the saturation of the UBHF energies is associated with the 
ability of the trapped bosons (independent of 
$N$) to minimize their mutual repulsion by occupying different positions in 
space, and this is one of our central results. 
For $N=2$, the two bosons localize at a distance $2a$ apart to form an 
antipodal dimer. For $N \leq 5$ the preferred UBHF crystalline arrangement is
a single ring with no boson at the center [usually denoted as $(0,N)$].
$N=6$ is the first case having one boson at the center [designated as 
$(1,N-1)$], and the (0,6) arrangement is a higher energy isomer.

The saturation found here for 2D trapped bosons interacting through 
strong repelling contact potentials is an illustration of the
``fermionization'' analogies that appear in strongly correlated systems
in all three dimensionalities. Indeed such energy saturation has been
shown for the TG 1D gas \cite{gir,gir2}, and has also been discussed for
certain 3D systems (i.e., three trapped bosons \cite{blum} and an infinite
boson gas \cite{heis}). Saturation of the energy
and the length of the trapped atom cloud (and thus of the interparticle
distance) has been measured recently for the 1D TG gas
(see in particular Fig.\ 3 and Fig.\ 4 in Ref.\ \cite{wei} and compare
to the similar trends predicted here for the 2D case in Fig.\ 1 and Fig.\ 2).

For the Coulomb potential [see Fig.\ 1(b)], 
the displayed total energies have been referenced to the classical energy 
$E^{\text{cl}}_C$  \cite{yl3} (plus the zero-point energy) 
of six trapped point charges in their (1,5) equilibrium configuration, since 
the total energy of a Wigner crystallite (independently of whether
it consists of bosons or fermions) is expected to approach
$E^{\text{cl}}_C$ as $R_W \rightarrow \infty$.
We see again that the $E_{\text{RBHF}}^{\text{Gauss}}$ energies
(one common Gaussian orbital) 
diverge as $R_W \rightarrow \infty$. In contrast, the unrestricted 
HF energies $E_{\text{UBHF}}$ remain finite and 
approach slowly $E^{\text{cl}}_C$ as $R_W \rightarrow \infty$. 
A similar behavior is exhibited by the total energies for 
all $N=2 - 7$ cases of charged bosons.

\begin{figure}[t]
\centering\includegraphics[width=7.2cm]{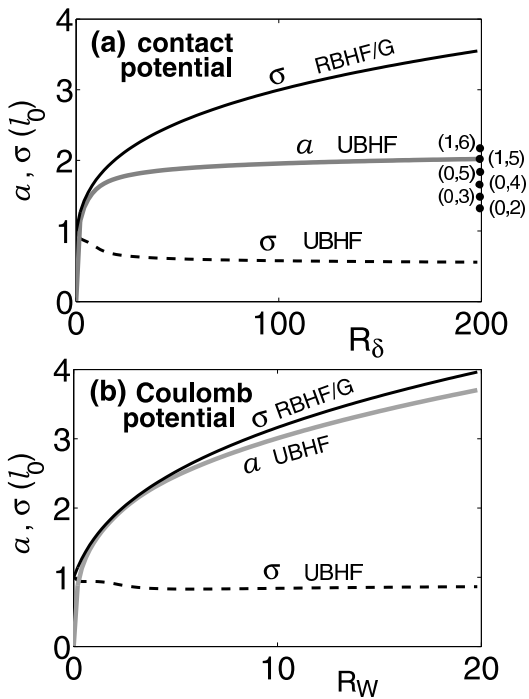}
\caption{
Variationally determined widths ($\sigma$) and ring radii ($a$) for $N=6$
harmonically confined 2D bosons as a function of (a) $R_\delta$ and (b) $R_W$, 
obtained according to the various approximations 
(as marked in the figure, see also caption of Fig.\ 1 ).
The saturation values of $a$ of the lowest-energy configuration
for $2 \leq N \leq 7$ are on the right in (a).
Lengths in units of $l_0$. For the UBHF case [displaying an (1,5) crystallite]
the interparticle distance on the pentagonal shell is $d= ((5-5^{1/2})/2)^{1/2}
a \approx 1.176 a$, showing the same saturation trend as the radius $a$. 
}
\end{figure}

In Fig.\ 2, we display for the $N=6$ bosons the radii of the polygonal rings 
$a$ and widths $\sigma$ of the Gaussian orbitals obtained in various 
approximations, as a function of $R_{\delta}$ (a) and $R_W$ (b). 
For the contact potential, in the RBHF/G approximation 
we find that $a=0$ and the width (marked as RBHF/G in Fig. 2)  
keeps increasing continuously as $R_{\delta} \rightarrow \infty$ 
(this reflects the unsuccessful attempt of the common orbital
to minimize the mutual repulsion between the bosons by spreading out as far
as possible). In contrast, the 
unrestricted widths $\sigma_{\text{UBHF}}$  associated with the displaced 
Gaussian orbitals (that correspond to a lower total energy, see Fig.\ 1) 
saturate to a constant value. Similar behaviors are also exhibited by 
$\sigma_{\text{RBHF}}^{\text{Gauss}}$ and $\sigma_{\text{UBHF}}$ in the case 
of a Coulomb force [see of Fig.\ 2(b)].

The radii $a$ associated with the pentagonal ring of localized orbitals,
however, exhibit a different behavior depending on whether the repulsive
potential is a contact or a Coulomb one. Indeed, in the Coulomb case, 
the radii $a_{\text{UBHF}}$ keep increasing with $R_W$, approaching the 
equilibrium radius $a^{\text{cl}}_C= 1.334 l_0 R_W^{1/3}$ of six $Ze$ classical
point-charges in a harmonic trap in the (1,5) configuration \cite{yl3}. 
In contrast, for a repulsive contact potential, the radii 
$a_{\text{UBHF}}$ saturate to a constant value $\approx 2 l_0$. 
The dependence of the saturation values of $a$ on $N$ (for $3 \leq N \leq 7$)
for the lowest-energy configurations is shown on the right in Fig.\ 2(a). The 
different behavior of the boson positions in the UHBF crystallite is a natural 
consequence of the long-range character of the Coulomb potential versus the 
short-range contact potential.

{\it Second step: Restoration of broken symmetry.\/} Although the optimized 
UBHF permanent $|\Phi_N \rangle$ performs exceptionally well regarding the 
total energies of the trapped bosons, in particular in comparison to the 
resctricted wave functions (e.g., the GP anzatz), it is still incomplete. 
Indeed, due to its localized orbitals, $|\Phi_N \rangle$ 
does not preserve the circular 
(rotational) symmetry of the 2D many-body hamiltonian $H$. Instead, it 
exhibits a
lower point-group symmetry, i.e., a $C_2$ symmetry for $N=2$ and a $C_5$ one
for $N=6$ (see below). As a result, $|\Phi_N \rangle$ does not have a good
total angular momentum. This is resolved through a
post-Hartree-Fock step of {\it restoration\/} of broken symmetries 
via projection techniques [13(b)],\cite{yl3}, yielding a new wave 
function $|\Psi_{N,L}^{\text{PRJ}} \rangle$ \cite{note45} with a
definite angular momentum $L$. Here, we focus on the properties of the ground 
state, i.e., $L=0$; the corresponding energy is $E_0^{\text{PRJ}}$. 

For $N=6$ 2D bosons, Fig.\ 1 shows that the $E_0^{\text{PRJ}}$ energies
share with the UBHF ones the saturation property for the case of a 
contact-potential  
repulsion, as well as the property of converging to $E^{\text{cl}}_C$
as $R_W \rightarrow \infty$ for the case of a Coulomb repulsion. In both
cases, however, the projections bring further lowering \cite{note43}
of the total energies
compared to the UBHF ones. Thus, for strong interactions 
(large values of $R_\delta$ or $R_W$) the restoration-of-broken-symmetry step 
yields an excellent approximation of both the exact many-body wave function and
the exact total energy \cite{note44}.

\begin{figure}[t]
\centering\includegraphics[width=8.0cm]{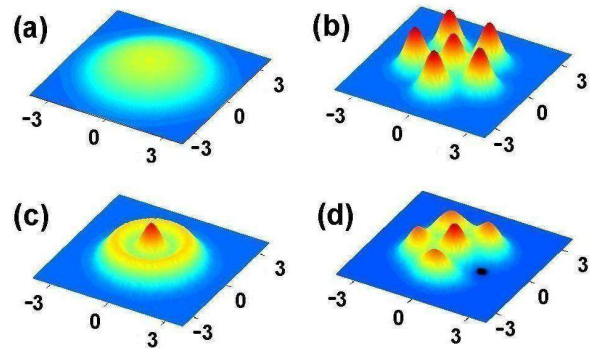}
\caption{
(a-c): Single-particle densities for $N=6$ 2D harmonically trapped neutral 
bosons with a contact interaction and $R_\delta=25$.
(a) The single-orbital self-consistent GP case. 
(b) The symmetry broken UBHF case (static crystallite).
(c) The projected (symmetry-restored wave function, see Ref.\ \cite{note45})
case (collectively fluctuating crystallite).  
The crystalline structure of the outer ring in this last case is ``hidden'',
but it is revealed in the conditional probability distribution \cite{yl4} 
displayed in (d), where the observation point is denoted by a black dot (on the right). 
Lengths in units of $l_0$.
}
\end{figure}

The transformations of the single-particle densities (displayed in Fig.\ 3
for $N=6$ neutral bosons interacting via a contact potential 
and $R_\delta=25$) 
obtained from application of the successive approximations provide an 
illustration of the two-step method of symmetry breaking with subsequent 
symmetry restoration. Indeed, the GP 
single-particle density [Fig.\ 3(a)] is circularly symmetric, but the UBHF one 
[Fig.\ 3(b)] explicitly exhibits a (1,5) crystalline configuration. After 
symmetry restoration [Fig.\ 3(c)], the circular symmetry is
re-established, but the single-particle density is radially modulated unlike 
the GP density. In addition, the crystalline structure in the projected wave 
function is now hidden; however, it can be revealed through the use of the
CPD \cite{yl4} [see Fig.\ 3(d)], which resembles the 
(crystalline) UBHF single-particle density, but with one of the humps 
on the outer ring missing (where the observer is located). In particular, 
$P({\bf r}_0, {\bf r}_0) \approx 0$ and the boson associated with the observer 
is surrounded by a ``hole'' similar to the exchange-correlation hole in 
electronic systems. This is another manifestation of the ``fermionization'' of
the strongly repelling 2D bosons. However, here as in the 1D TG case 
\cite{gir,gir2}, the vanishing of $P({\bf r}_0,{\bf r}_0)$ results from the
impenetrability of the bosons. For the GP condensate, the CPD 
is independent of ${\bf r}_0$, i.e., 
$P_{\text{GP}}({\bf r}, {\bf r}_0) \propto |\phi_0({\bf r})|^2$, reflecting
the absence of any space correlations.

It is of importance to observe that the radius of the BEC [GP case, Fig.\ 3(a)] 
is significantly larger than the actual radius of the 
strongly-interacting crystalline phase [projected wave function, Fig.\ 3(c)]. 
This is because the extent of the crystalline phase 
saturates, while that of the GP condensate grows with no bounds as $R_\delta
\rightarrow \infty$. Such dissimilarity in size (between the condensate and 
the strongly-interacting phase) has been also predicted \cite{dunj} for the 
trapped 1D Tonks-Girardeau gas and indeed observed experimentally \cite{wei}. 
In addition, the 2D single-particle momentum distributions for neutral bosons 
have a one-hump shape with a maximum at the origin (a behavior exhibited also 
by the trapped 1D TG gas). The width of these momentum distributions versus 
$R_\delta$ increases and saturates to a finite value, while that
of the GP solution vanishes as $R_\delta \rightarrow \infty$.

In conclusion, we provided a solution to strongly repelling bosons in 2D 
harmonic traps using a two-step method of breaking of rotational symmetry at 
the unrestricted Bose-Hartree-Fock level and of subsequent symmetry 
restoration. This method yields substantially lower total energies 
compared to the GP solution, through the inclusion of correlations beyond  
the single-orbital Bose-Einstein condensate. We find that the bosons become 
localized and form crystalline patterns made of concentric polygonal rings, 
both for a repulsive contact and a Coulomb interaction. For neutral bosons 
the total energy of the crystalline phase saturates with increasing strength 
of the repulsion, in contrast to the GP condensate whose energy diverges. 
Furthermore, the spatial extent saturates and becomes smaller than that of the 
GP condensate, which grows without limit. For charged bosons, the total energy 
and spatial extent of the crystalline phase approach the classical values of 
point-like charges in their equilibrium configuration as 
$R_W \rightarrow \infty$. In light of the above, we trust that our 
predictions will provide the impetus for experimental efforts to access the
regime of strongly repelling bosons in two dimensions. To this end we 
anticipate
that extensions of methodologies developed for the recent realization of the 
Tonks-Girardeau regime in 1D (using a finite small number of trapped $^{87}$Rb 
and optical lattices, with a demonstrated wide variation of $R_{\delta}$ 
from 5 to 200 \cite{par} and from 1 to 5 \cite{wei}) will prove most promising.
Control of the interaction strength via the use of the Feshbach resonance
may also be considered \cite{cor}.

This research is supported by the U.S. D.O.E. (Grant No. FG05-86ER45234 ) and 
by the NSF.


\begin{thebibliography}{99}
\bibitem{note2}
For strongly correlated electrons in parabolic quantum dots the 
Gaussian form is adequate in most cases, leading to
formation of electron crystallites \cite{yl} (Wigner molecules).
\bibitem{note45}
The projected multi-permanent wave function can be written as
$ 2 \pi |\Psi_{N,L}^{\text{PRJ}} \rangle = \int^{2\pi}_0 d\gamma
|\Phi_{N}(\gamma) \rangle \exp(i\gamma L)$,
where $|\Phi_{N}(\gamma) \rangle$ is the original UBHF permanent having each 
localized orbital rotated by an azimuthal angle $\gamma$, with $L$ being  
the total angular momentum. The projection yields wave functions for a whole 
rotational band. The projected ground-state ($L=0$) energy is given by 
$E_0^{\text{PRJ}} = \langle \Psi_{N,0}^{\text{PRJ}}  | H | 
\Psi_{N,0}^{\text{PRJ}}  \rangle / 
\langle \Psi_{N,0}^{\text{PRJ}}  | \Psi_{N,0}^{\text{PRJ}}
\rangle$.
\end{thebibliography}
\end{document}